\documentclass[aip, rsi, amsmath, amssymb, reprint]{revtex4-1}

\usepackage{hyperref}

\usepackage{graphicx}
\usepackage{siunitx}
\usepackage{gensymb}

\begin{document}

\title{Position Sensitive Alpha Detector for an Associate Particle Imaging System}

\author{Mauricio Ayllon Unzueta}
\email{mayllonu@lbl.gov}
\affiliation{Accelerator Technology \& Applied Physics Division, E.O. Lawrence Berkeley National Laboratory, Berkeley, CA 94720, USA}
\author{Will Mixter}
\affiliation{Accelerator Technology \& Applied Physics Division, E.O. Lawrence Berkeley National Laboratory, Berkeley, CA 94720, USA}
\author{Zachary Croft}
\affiliation{Accelerator Technology \& Applied Physics Division, E.O. Lawrence Berkeley National Laboratory, Berkeley, CA 94720, USA}
\author{John Joseph}
\affiliation{Engineering Division, E.O. Lawrence Berkeley National Laboratory, Berkeley, CA 94720, USA}
\author{Bernhard Ludewigt}
\affiliation{Accelerator Technology \& Applied Physics Division, E.O. Lawrence Berkeley National Laboratory, Berkeley, CA 94720, USA}
\author{Arun Persaud}
\email{apersaud@lbl.gov}
\affiliation{Accelerator Technology \& Applied Physics Division, E.O. Lawrence Berkeley National Laboratory, Berkeley, CA 94720, USA}

\begin{abstract}
Associated Particle Imaging (API) is a nuclear technique that allows for the nondestructive determination of 3D isotopic distributions. The technique is based on the detection of the alpha particles associated with the neutron emitted in the deuterium-tritium (DT) fusion reaction, which provides information regarding the direction and time of the emitted \SI{14}{\MeV} neutron. Inelastic neutron scattering leads to characteristic gamma-ray emission from certain isotopes, for example $^{12}$C, that can be correlated with the neutron interaction location. An API system consisting of a sealed-type neutron generator, gamma detectors, and a position-sensitive alpha detector is under development for the nondestructive quantification of carbon distribution in soils. This paper describes the design of the alpha detector, detector response simulations, and first experimental results. The alpha detector consists of a Yttrium Aluminum Perovskite (YAP) scintillator mounted on the inside of a neutron generator tube. The scintillation light propagates through a sapphire window to a position-sensitive photomultiplier tube mounted on the outside. The $16 \times 16$ output signals are connected through a resistive network for a 4-corner readout. The four readout channels are amplified, filtered, and then digitized for the calculation of the alpha position. First test results demonstrate that an excellent alpha position resolution, better than the \SI{1}{\mm} FWHM required by the application, can be achieved with this detector design.
\end{abstract}

\maketitle

\section{INTRODUCTION}
\begin{figure}[htbp]
  \centerline{\includegraphics[width=0.5\textwidth]{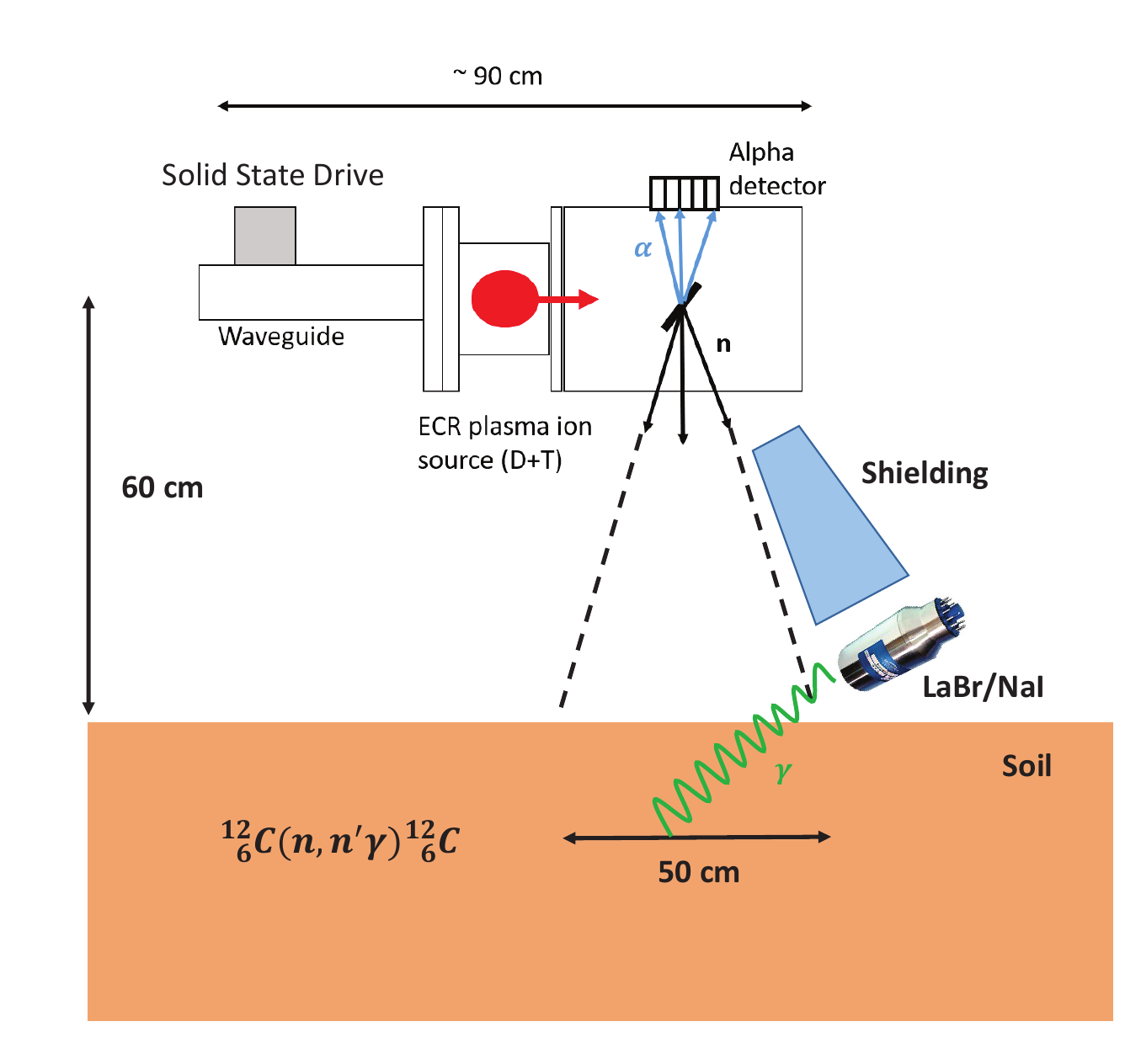}}
  \caption{Simplified experimental setup showing the main components of an associated particle imaging system and the proposed geometry to measure carbon concentration in soil.}
  \label{Fig:setup}
\end{figure}

The concentration of carbon in soil is an important parameter that relates to soil health and
productivity. Improved land management practices and agricultural plants have the potential to increase carbon in soil and thus contribute to carbon sequestration.
As part of an ARPA-e funded effort to improve soil health evaluation by creating new technologies and sensors, we are developing an instrument to directly and non-destructively measure the carbon concentration and distribution in soil under ARPA-e's ROOTS program \cite{ROOTS}. Our approach is based on measuring gamma rays from neutron inelastic scattering events. The experimental setup is shown in Figure \ref{Fig:setup}.
The instrument consists of a deuterium-tritium (DT) neutron generator, an alpha particle
detector, and gamma-ray detectors. The DT neutron generator (dt108api, Adelphi Technology \cite{Adelphi}) is located approximately \SI{60}{cm} above the soil. It generates neutrons via DT fusion reactions as shown in Equation \ref{eq:reactionDT}.
\begin{equation}\label{eq:reactionDT}
    D + T \rightarrow \alpha\ (\SI{3.5}{\MeV}) + n\ (\SI{14}{\MeV})
\end{equation}
In this reaction, an alpha particle is emitted with each neutron in approximately opposite directions as dictated by the reaction kinematics. The alpha particle can then be detected with a position-sensitive detector and the detection time recorded. This system allows for the calculation of the direction of the
time-tagged neutrons. If a tagged neutron induces an inelastic scattering
reaction on a $^{12}$C nucleus located in the soil, a \SI{4.4}{\MeV} gamma ray is emitted from the transition of the first excited state of $^{12}$C to the ground state. Gamma-ray detectors (in our proof of principle system LaBr$_3$ and NaI detectors) are located close to the inspected soil. If a gamma ray is detected in coincidence with the alpha particle, the location of the emitting nucleus can be calculated from the direction of the neutron and the arrival time difference between the alpha and the gamma-ray. Accumulating many events allows for the determination of the 3D carbon distribution in a region of \SI{50}{\cm} $\times$ \SI{50}{\cm} $\times$ \SI{30}{\cm (depth)} of soil. Simulations show that
by using this method we can achieve an overall spatial resolution of several centimeters provided the position and time resolutions of the alpha detector are sufficient. Measurement times may vary from a few minutes to hours depending on the desired measurement precision, and depth of soil layer inspected. In this paper, we report on the alpha detector design, first measurements of its position resolution using a radioactive source, and discuss aspects of its influence on the overall performance of the instrument.

\section{ALPHA DETECTOR DESIGN}
The purpose of the pixelated alpha detector is twofold: it is used to obtain the starting time for time-of-flight measurements (depth resolution), and it allows for the calculation of the neutron direction of travel (x-y resolution). Our goal is to measure carbon distribution in soil with a voxel resolution of \SI{5}{\cm}. Given that a \SI{14}{\MeV} neutron travels approximately \SI{5}{\cm} in \SI{1}{\ns}, the system needs to have a time resolution better than \SI{1}{\ns}. Additionally, in order to achieve a x-y resolution in the soil of approximately \SI{5}{cm}, the alpha detector needs to have a position resolution of approximately \SI{1}{mm} given the geometry shown in Figure \ref{Fig:setup}. However, these two requirements also depend on other system parameters such as the time resolution of the gamma detectors and the neutron beam spot size, respectively. The alpha detector should also be able to handle high alpha rates since we expect approximately $\SI{1e7}{alphas\per\second}$ at the maximum neutron output of the neutron generator (\SI{2e8}{neutrons\per\second}). Because the alpha detector needs to be placed inside the sealed-type neutron generator, the detector material has to withstand bake-out temperatures of over \SI{100}{\degreeCelsius}.
Therefore, the choice is limited to inorganic scintillators, which have a fast response and short decay times, decent light yield,  and sufficient energy resolution.
The inorganic scintillator Yttrium Aluminum Perovskite (YAP) is used since its rise time is a few hundred picoseconds and its decay time is approximately \SI{27}{\ns} \cite{Moszynski1998}. Additionally, the light yield is around \SI{17000}{photons\per\MeV} for gamma rays ($\sim\!3$ times lower for alphas) \cite{Wolszczak2017}, and it is non-hygroscopic; hence easy to handle.

The dimensions of the YAP crystal are
$\SI{50}{\mm}\times\SI{50}{\mm}\times\SI{1}{\mm}$ (manufactured by Crytor \cite{Crytur}) and it is placed approximately
\SI{6}{\cm} from the titanium target where the neutrons and alpha particles are created.

As the alpha particles enter the YAP crystal, they lose energy mainly due to collisions with electrons and thereby transfer their energy to the crystal. All alpha particles will stop
within several micrometers inside the YAP. This process creates a large number of electron-hole pairs, which upon de-excitation through activator sites (cerium) emit optical photons with a primary wavelength of approximately \SI{370}{\nm} \cite{Knoll}. This wavelength is well-matched to our PMT spectral sensitivity.
In the YAP crystal, around \SI{6000}{photons\per\MeV} are emitted in a single \SI{3.5}{\MeV} alpha particle interaction \cite{Moszynski1998}. The photons then
propagate through a \SI{3}{mm}-thick vacuum sapphire window (MPF Products \cite{MPF}) to the photocathode of the PMT
(Hamamatsu H13700-03 \cite{Hamamatsu}). To further increase the number of photons reaching the PMT, the YAP surface facing the alpha source was coated with a \SI{400}{\nm} thick aluminum layer functioning as a mirror. Furthermore, the metal layer serves as a ground plane to prevent
the scintillator from charging up due to the deposition of positively charged alpha particles. During final
integration of the system with the neutron generator, as shown in Figure \ref{Fig:setup}, an additional aluminum foil of \SI{1}{\um} thickness
is used in between the titanium target and the YAP crystal in order to provide a pinhole free layer to protect the photomultiplier from ion source light.
A layer of a certain thickness (for aluminum $\SI{>1}{\um}$) is also needed to prevent scattered
deuterium and tritium ions from the ion beam to reach the scintillator. The foil together with the existing coating of the YAP provides the required total thickness. The energy loss of the alpha particle in the foil is of the order of \SI{50}{\keV} and is therefore negligible.
Sapphire was chosen as the vacuum window material for its superior optical properties and mechanical strength,
which allows for a significant reduction of the required thickness of the window, and therefore reduces
the spread of light from a single alpha-particle interaction onto the photocathode. This is in contrast to several other recent designs that use a fiber optic faceplate to couple the scintillator light to the PMT, such as in \cite{Zhang2012}.

The PMT is operated at a voltage bias of \SI{-1}{\kV} and the anode is pixelated into an array of $16\times16$ individual electrodes, each of them with an active area of $\SI{2.8}{\mm}\times\SI{2.8}{\mm}$ and \SI{3}{\mm} center-to-center distance. The signals can be read out from those individual pixels as well as from the penultimate dynode, which serves as a common timing signal for fast timing applications.

\section{ALPHA DETECTOR SIMULATIONS}
\begin{figure}[htbp]
  \centerline{\includegraphics[width=\linewidth]{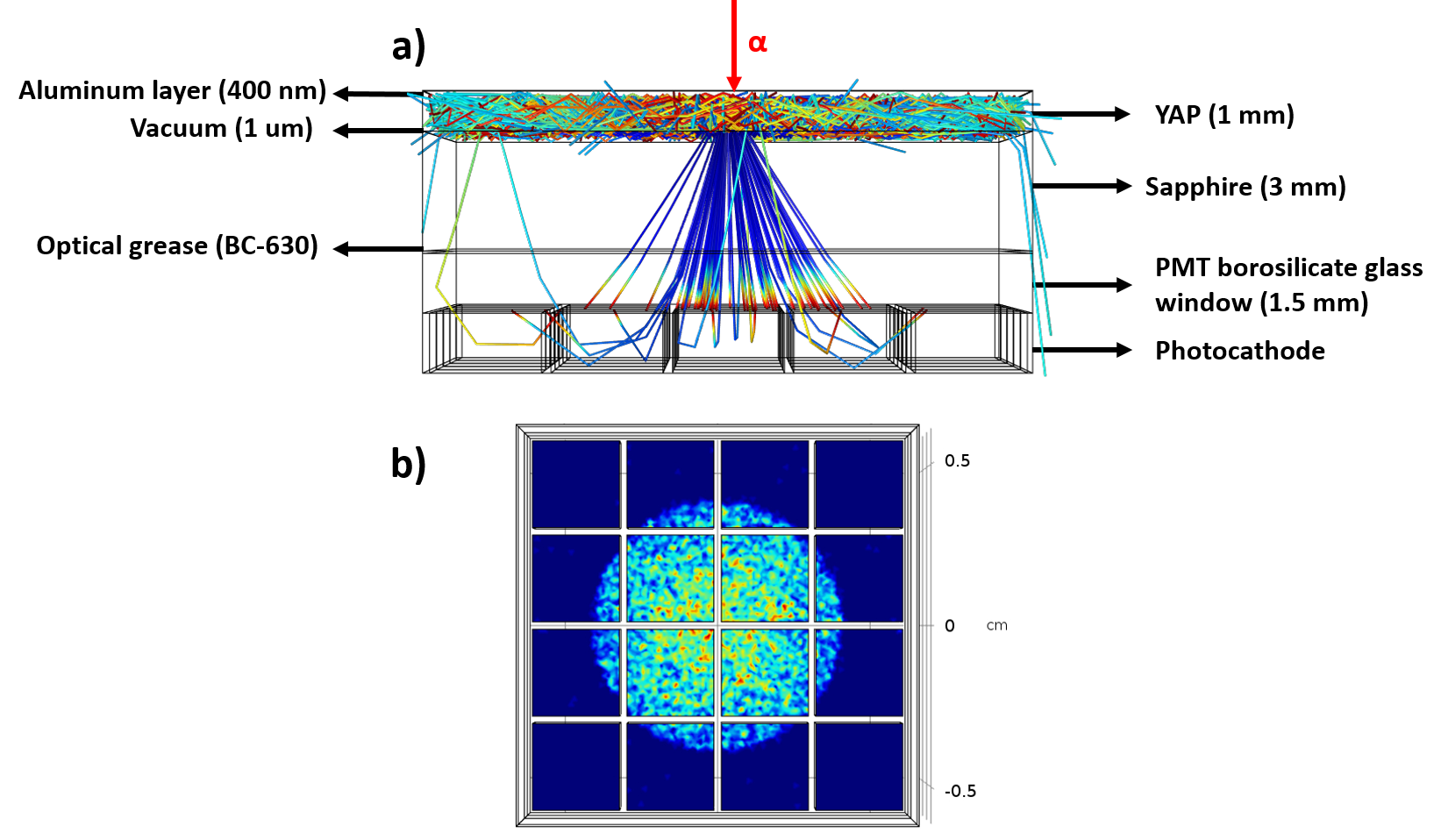}}
  \caption{Comsol Multiphysics \cite{Comsol} simulation of optical photon transport through the alpha detector system showing a) the simulation geometry (side view) and ray tracing for a single alpha event, and b) the light spread (top view) onto the photocathode for a single alpha interaction. The pixel size  of the detector is \SI{2.8}{\mm} and the pitch is \SI{3.04}{\mm}.}
  \label{Fig:comsol-alpha-det}
\end{figure}

The finite-element analysis code Comsol Multiphysics \cite{Comsol} was used to simulate the optical photon transport through the scintillator and vacuum window and up to the photocathode of the PMT. These simulations are important in order to optimize the geometry and estimate the amount of photoelectrons produced per event, which directly influences the shape of the signal. They also help us understand the light spread onto the photocathode, which influences the choice of reconstruction algorithm, and hence the ultimate attainable position resolution. Figure \ref{Fig:comsol-alpha-det} a) shows that most photons from an alpha particle interaction do not reach the sapphire window due to total internal reflection at the scintillator-vacuum interface because of the mismatch of their indices of refraction. Unfortunately, due to the requirements of the neutron generator it is not possible to use optical grease to better match the indices of refraction at this interface. Simulations show that for a \SI{3.5}{\MeV} alpha particle we can only expect approximately 600 photoelectrons in the first stage of the dynode structure of the PMT, even though a total of approximately 21000 photons are created in the YAP crystal per alpha event. These simulations further show that light spreads approximately over a $3\times3$ array of pixels, as shown in Figure \ref{Fig:comsol-alpha-det} b).

The fact that there is a finite light spread onto the photocathode means the position resolution can be even smaller than the pixel size depending on what reconstruction algorithm is used e.g. center of mass reconstruction, Anger logic, etc. For example, J. W. Cates et al. \cite{Cates2013} demonstrated a position resolution of $<$ \SI{0.5}{\mm} with a similar experimental setup and using a 4-corner readout scheme.

In order to quantify the influence of the alpha detector on the overall system performance, MCNP (version 6.1) \cite{MCNP}
simulations were performed using the ``ptrac'' function, which records event by event neutron and gamma ray histories. The simulation consists of a \SI{14.1}{\MeV} neutron point source placed \SI{60}{\cm}
above the soil with a density of \SI{1.3}{\g\per\cm\cubed} and a mixture of 51\% oxygen, 35\% silicon, 8\% aluminum, 4\% iron, and 2\% carbon (atom percent). Additionally, a block of graphite is placed inside
the homogeneous soil, \SI{10}{\cm} below the surface. A single 5-inch NaI gamma detector was placed above the surface outside the API cone.

\begin{figure}[htbp]
  \centerline{\includegraphics[width=0.7\linewidth]{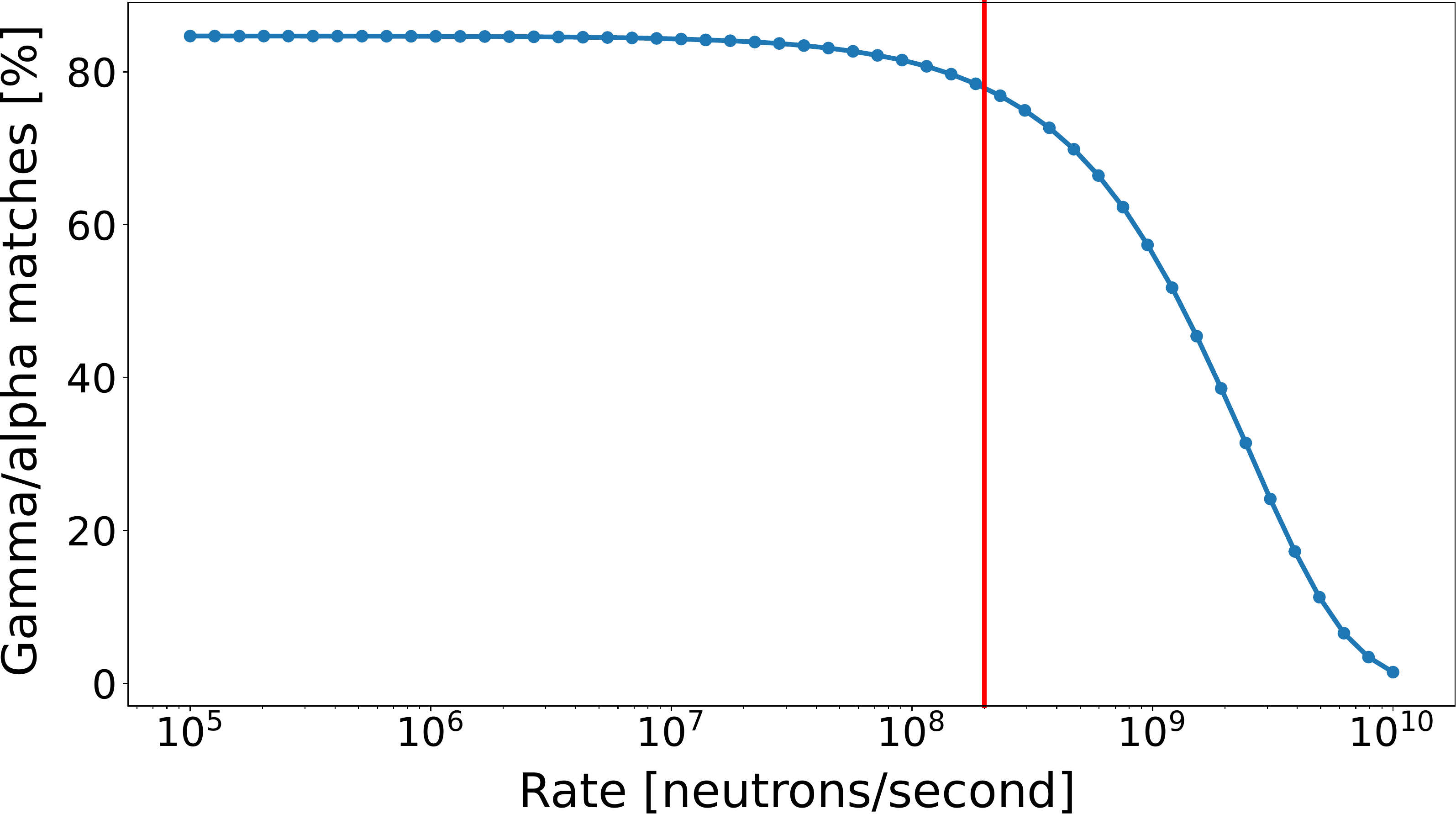}}
  \caption{ Results of an MCNP simulation showing the effect of the neutron rate on the percentage of unique alpha-gamma pairs detected. The red line indicates the maximum neutron yield of our generator at \SI{2e8}{neutrons\per\second}}
  \label{Fig:rate}
\end{figure}

The position of the alpha interaction in the detector was calculated and recorded together with the arrival time of the gamma ray. With this information, we reconstructed the position of the neutron interaction with carbon nuclei in the soil. Included in these events are multiple neutron and gamma-ray scattering events, which result in position errors exceeding \SI{5}{\cm} which are on the order of 5\% of the total number of events. The total number of neutrons simulated was \SI{6e10}{}, which is the equivalent of running our neutron generator for five minutes. The simulation is important in order to investigate two important aspects of the experiment: the influence of the position resolution of the alpha detector on the reconstruction of the carbon location and the effects of the neutron rate on the overall performance. For the latter, we assigned random uniformly-distributed times to the simulated neutrons within a five minute time window and examine coincidences between alphas and gamma rays. In order to correlate a gamma ray to an alpha particle, there needs to be a single coincident pair within a narrow window of time (we used \SIrange{7}{16}{\ns} after the arrival of the alpha particle). Since the rate of alpha particles arriving at the detector is higher than that of gamma rays, the main source of error stems from two or more alpha particles arriving within a short time. Another source of error is due to background gamma rays or multiple gamma rays arriving within a short time window. Events that cannot be assigned a single alpha to a gamma ray were discarded. Other events, such as a background gamma rays arriving within the expected time window of an alpha, give rise to an error in the reconstructed location of a given nucleus in the soil.
Figure \ref{Fig:rate} shows the result of the simulation mentioned above where the percentage of unique alpha-gamma pairs is plotted as a function of neutron rate. Note that for neutron rates higher than approximately $3-4\times 10^8$, there is a sharp drop-off in usable alpha-gamma pairs due to pile-up in the alpha detector. The reason why the percentage of usable alpha-gamma pairs do not reach 100\% at low rates is due to the fact that we limit the possible arrival time of the gamma ray that corresponds to the neutron time-of-flight for a soil depth of about \SI{30}{\cm}. Events from deeper soil layers are ignored in the analysis.

\begin{figure}[htbp]
  \centerline{\includegraphics[width=0.7\linewidth]{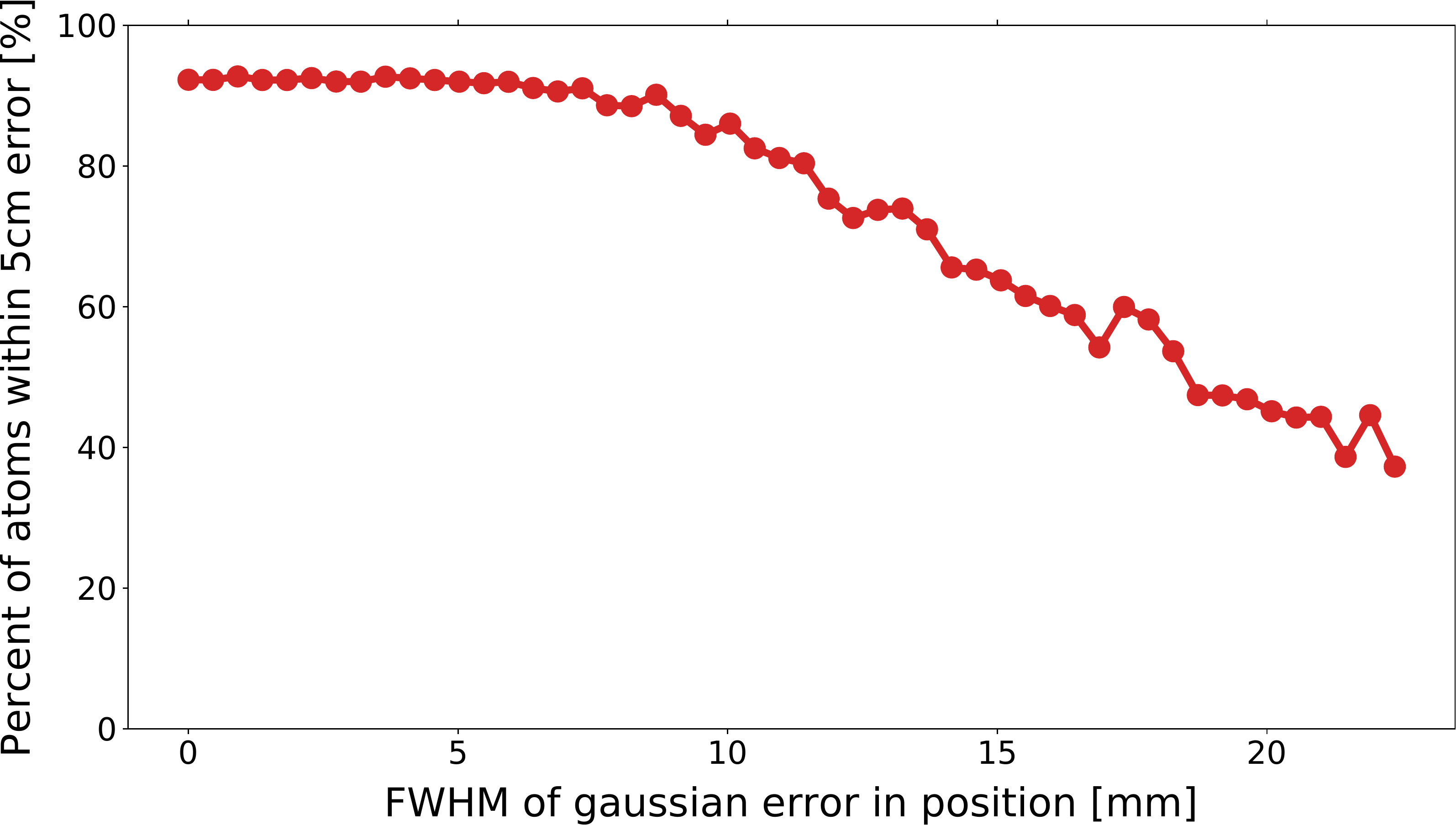}}
  \caption{Results of an MCNP simulation showing the influence on the carbon nuclei position resolution as a function of the position resolution of the detected alpha particle.}
  \label{Fig:gausserror}
\end{figure}

In order to quantify the carbon position errors stemming from the alpha position resolution, the same simulation is used to compare the calculated position of a carbon nucleus in the soil with the exact position known from the simulation. Figure \ref{Fig:gausserror} shows the percentage of total events that result in a position error or carbon nuclei smaller than $\SI{5}{\cm}$ as a function of the error in the calculated position of the alpha particle interaction in the detector. This error was assumed to be Gaussian, as shown in the Comsol Multiphysics simulations.
Since the initial distribution of errors is smaller than \SI{5}{\cm}, Figure \ref{Fig:gausserror} shows a flat response for small errors in the alpha position. For errors greater than approximately \SI{6}{\mm} FWHM,  the overall position resolution starts dropping.

The Monte Carlo simulation shows that larger errors can be accommodated in regard to the alpha detector position resolution. However, the main source of error in the overall system is projected to be caused by the neutron beam spot size.

\section{PMT Readout}\label{sec:readout}

\begin{figure}[htbp]
  \centerline{\includegraphics[width=\linewidth]{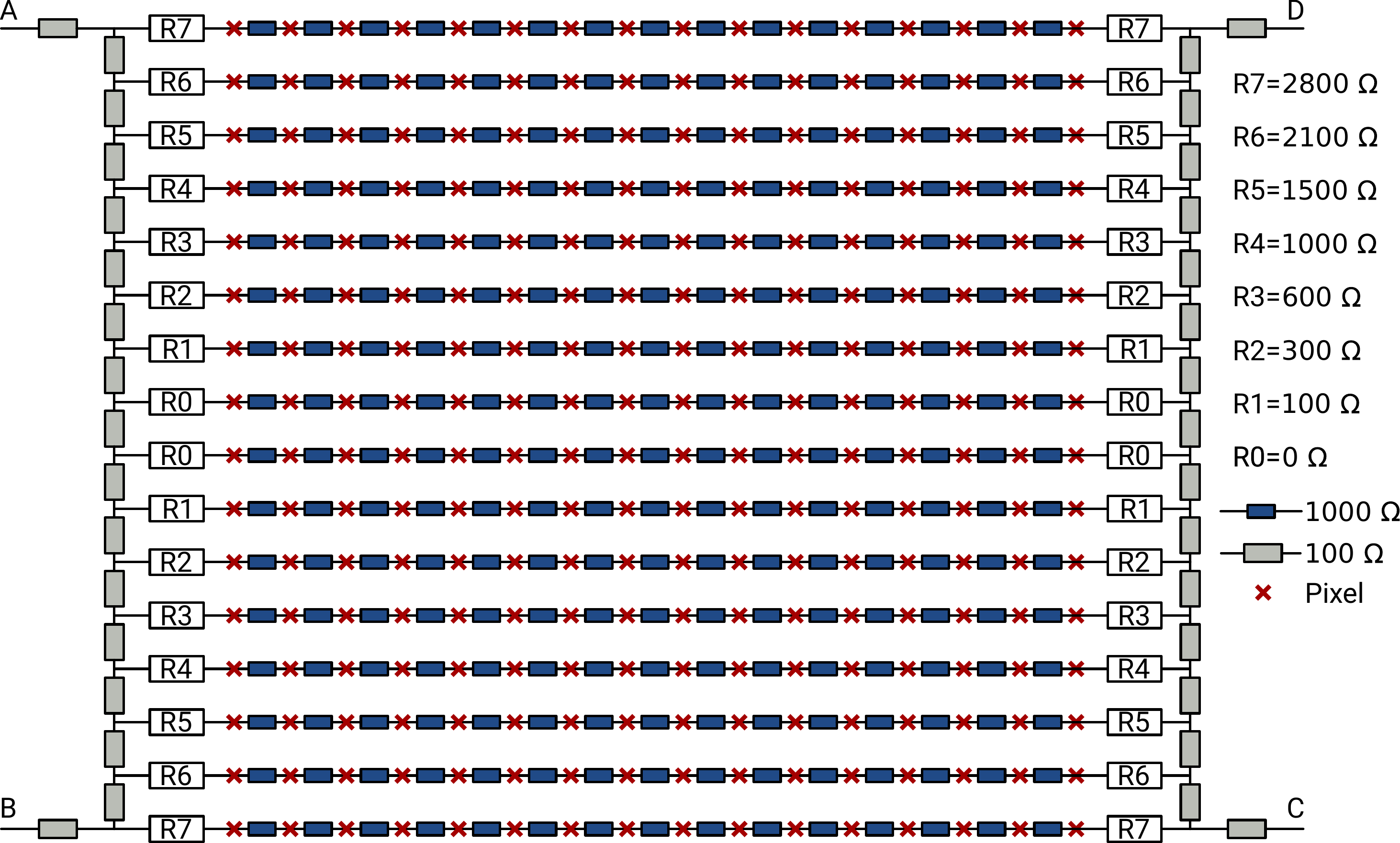}}
  \caption{A schematic of the 4-corner readout board. Rows of pixels are connected with \SI{1}{\kilo\ohm} resistors (blue) and terminated with a varying resistor at the end of each row. The rows are then connected using \SI{100}{\ohm} resistors (grey). The resulting four corners are connected to a pre-amplifier circuit and then to the digitizer.}
  \label{Fig:4cornerschematic}
\end{figure}

Several options exist for the PMT electronic readout. The 256 pixels can be all connected using a resistive network where only four corners are read out as shown in Figure \ref{Fig:4cornerschematic}. This is the more traditional method implemented in systems such as in \cite{Siegel1996}. However, the high alpha rates expected can be a limitation for this scheme. In order to overcome this, another option is to connect the rows and columns into single measurements units such as in \cite{Popov2006}, which decouples each anode output into two directions (X and Y). Therefore, higher rates can be handled, and if loading resistors are added, a gain correction per pixel can be implemented. However, more digitizers are needed (a total of 32), and the board layout becomes more complex given the large increase in number of resistors and the higher chances of cross-talk. Finally, the highest signal rate could be handled by digitizing every pixel. However, the data rates to be analyzed become large, and 256-channel digitizers would probably have to be implemented in a custom field-programmable gate array (FPGA), which would increase costs and development time significantly.

\begin{figure}[htbp]
  \centerline{\includegraphics[width=\linewidth]{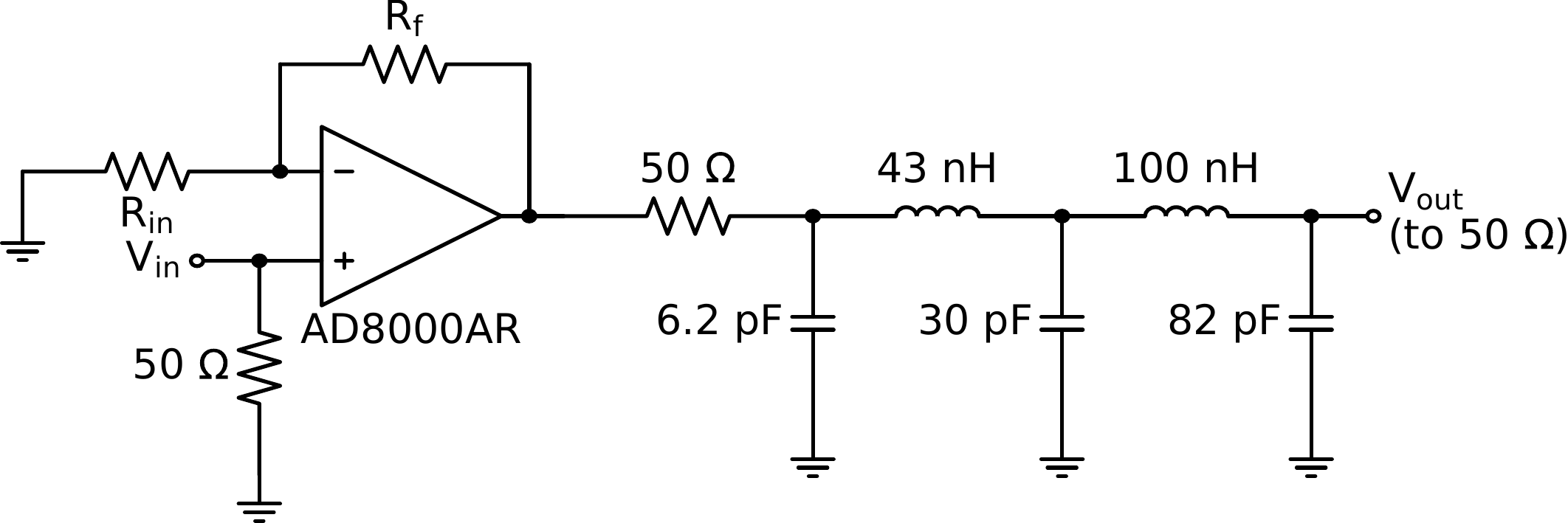}}
  \caption{The pre-amplifier design from XIA used for $\times10$ amplification as well as filtering of the signal for optimal digitization in the Pixie-16 system.}
  \label{Fig:preamp}
\end{figure}

We decided to first implement a 4-corner readout scheme to understand its rate limitations and position resolution. The approach was based on \cite{Siegel1996} and \cite{Olcott2003}. The resistive network was analyzed using the software package LTspice \cite{LTSpice} and designed to minimize distortion. For this purpose, the side resistor values shown in Figure~\ref{Fig:4cornerschematic} were chosen in order to provide a linear response with respect to interaction location.
The PMT signals were appropriately shaped by XIA-designed pre-amplifier circuits with $\times10$ gain and a low-pass filter to enable optimal arrival time detection in a fully digital Pixie-16 system (\SI{500}{\MHz}) from XIA \cite{XIA}. The pre-amplifier schematic is shown in Figure \ref{Fig:preamp}.

\begin{figure}[htbp]
  \centerline{\includegraphics[width=\linewidth]{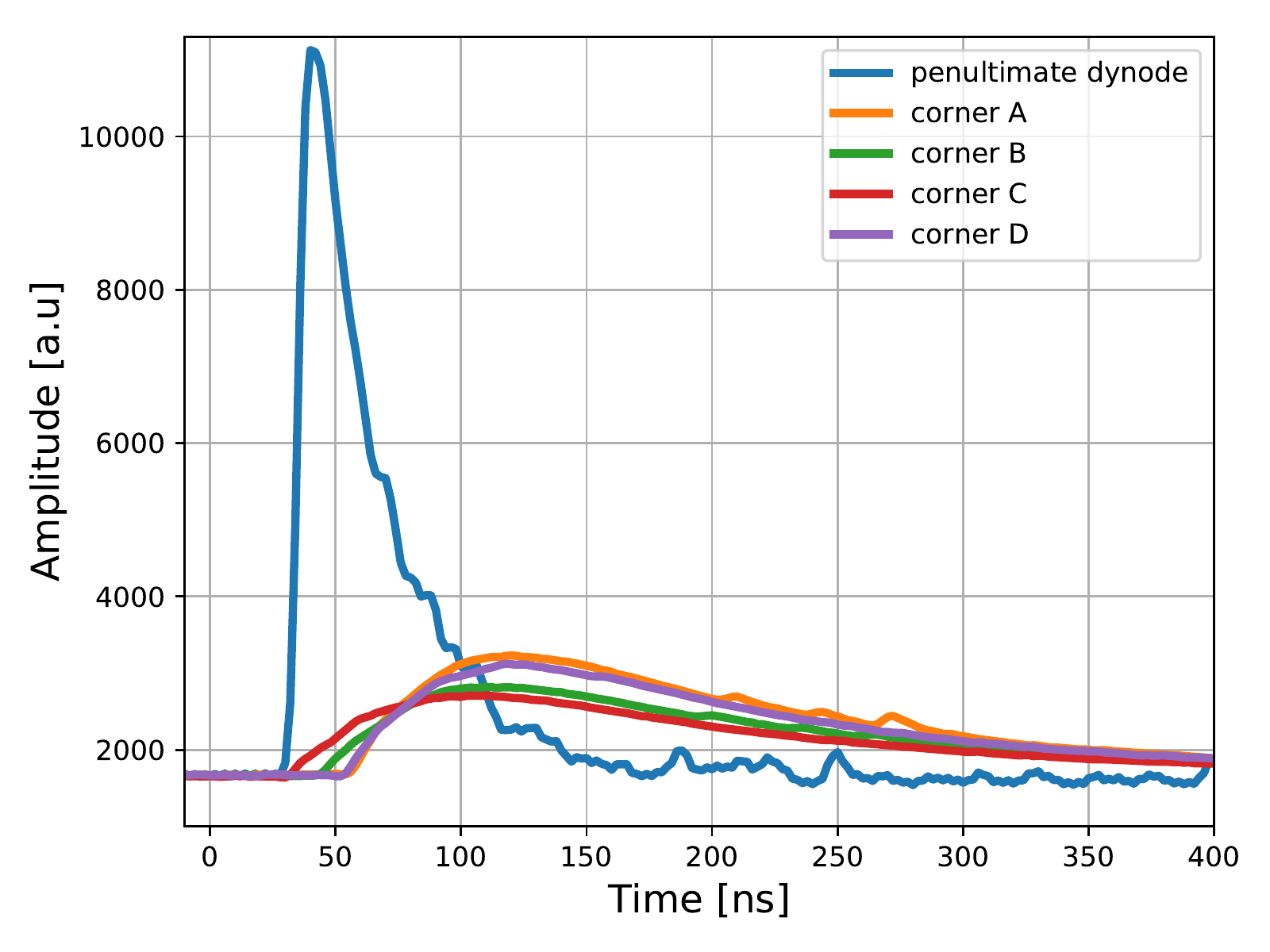}}
  \caption{Digitized traces from the last dynode of the PMT and from the 4-corner readout board. The traces from the four corners have a lower amplitude due the fact that the signal is distributed between the four corners and their rise time is lower due to additional RC delays from the resistive network.}
  \label{Fig:traces-4corner}
\end{figure}

\begin{figure}[htbp]
  \centerline{\includegraphics[width=0.8\linewidth]{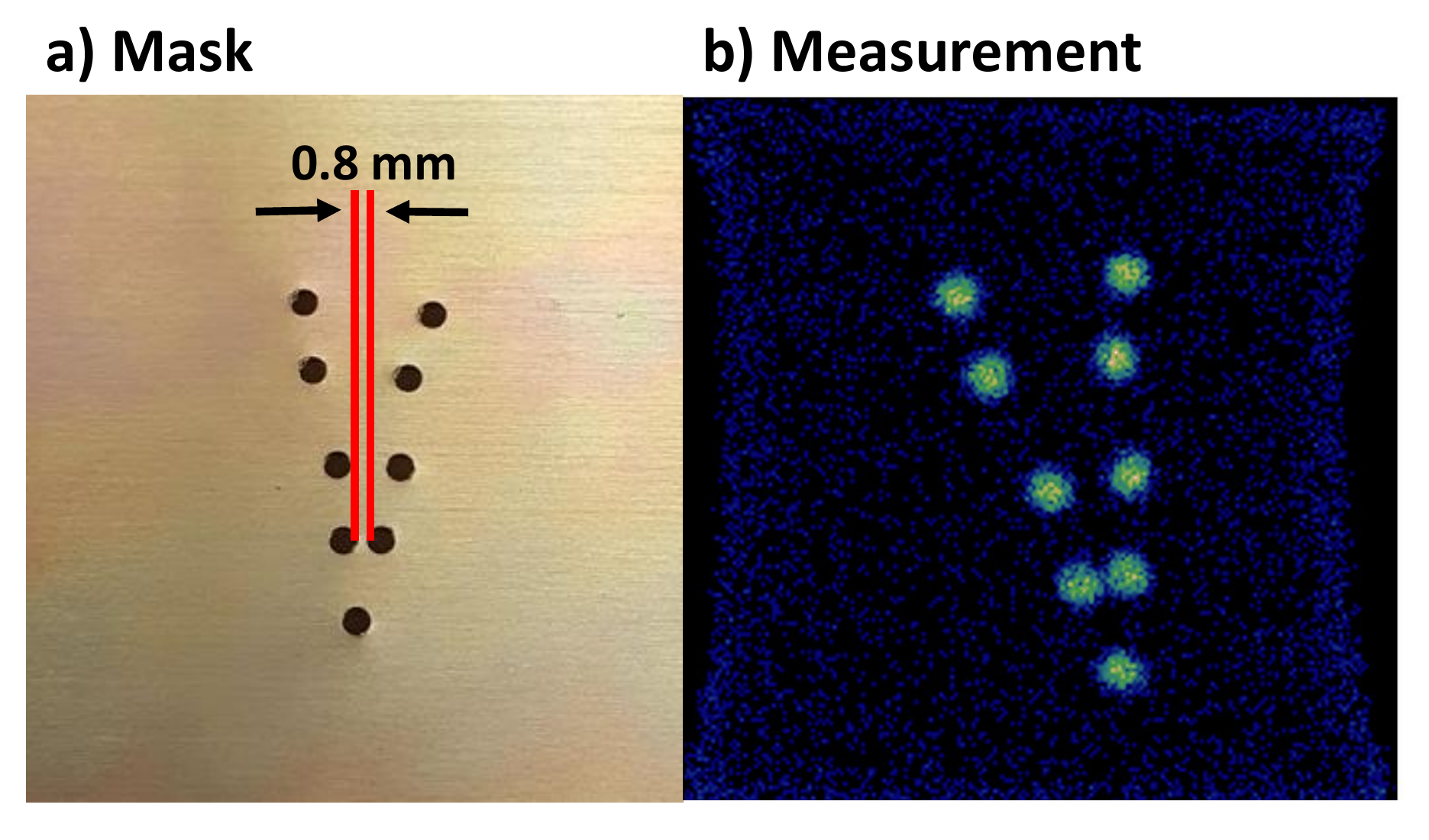}}
\caption{Image of a mask with a set of holes with varying distances among them. The \SI{0.8}{\mm} distance between the closest pair was clearly resolved.}
  \label{Fig:v-shape}
\end{figure}

The Pixie-16 was set up to trigger on the last common dynode (timing channel) of the PMT (the signal from the four corners has slower rise times due to the resistive network) and for each trigger the traces from each corner were automatically recorded. The position of interaction in the scintillator was then calculated using Equation~\ref{eq:xy}.
\begin{eqnarray}
x = \frac{A+B}{A+B+C+D}, \quad y = \frac{A+C}{A+B+C+D} \label{eq:xy}
\end{eqnarray}
where $A, B, C, D$ are the measured peak-to-peak values of the signals at the four corners.
Typical digitized waveforms of such an alpha interaction in YAP is shown in Figure \ref{Fig:traces-4corner}.
In order to quantify the position resolution of the alpha detector, masks with different hole patterns were placed in between a Pu-239 (\SI{900}{\becquerel}) alpha source and the YAP crystal in a 6-inch cube vacuum chamber. Pu-239 decays primarily by the emission of an alpha particle with an average energy of \SI{5.2}{\MeV} \cite{Browne2014}. Even though the energy of the alpha particles emitted by Pu-239 is higher than the \SI{3.5}{\MeV} alphas from the DT reaction, it bears no significance in terms of position resolution.  Figure \ref{Fig:v-shape} shows a mask with a series of hole pairs with varying distances between them.
The shortest distance between hole edges is \SI{0.8}{\mm}, and our system can clearly resolve them, hence showing a position resolution better than the required (\SI{1}{\mm}).

\begin{figure}[htbp]
  \centerline{
  \vtop{\vskip5pt\hbox{\includegraphics[width=0.33\linewidth]{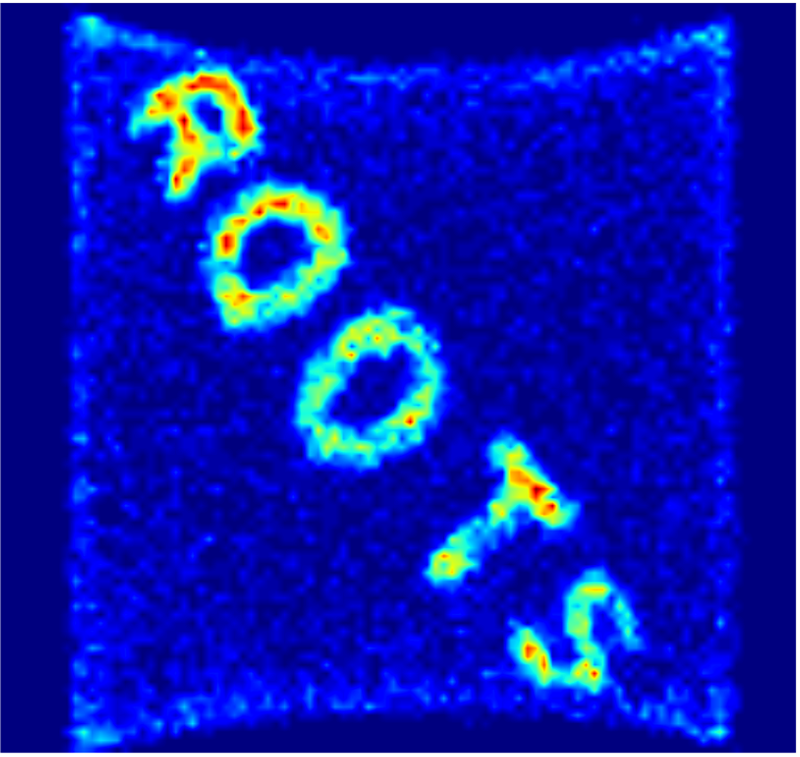}}}\hfill
  \vtop{\vskip0pt\hbox{\includegraphics[width=0.33\linewidth]{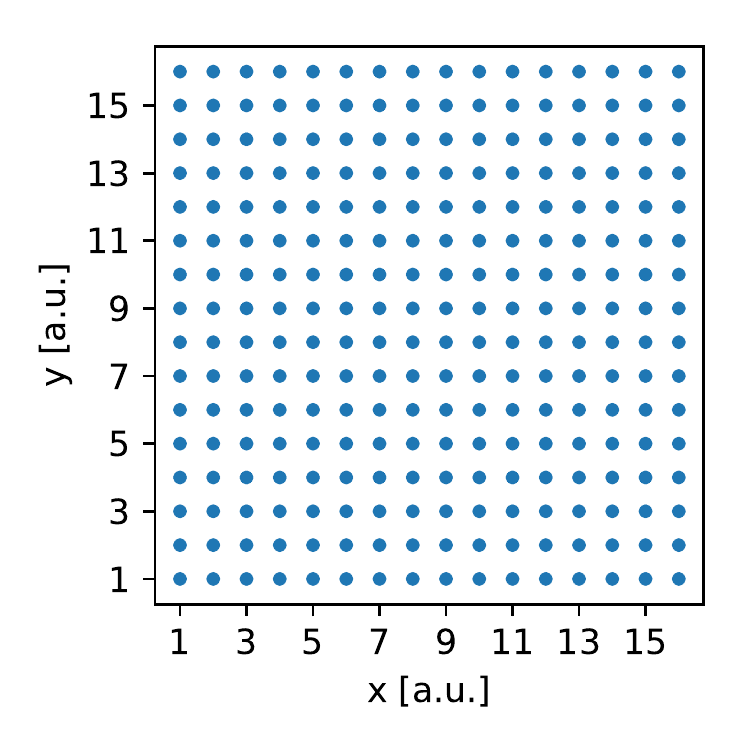}}}\hfill
  \vtop{\vskip0pt\hbox{\includegraphics[width=0.33\linewidth]{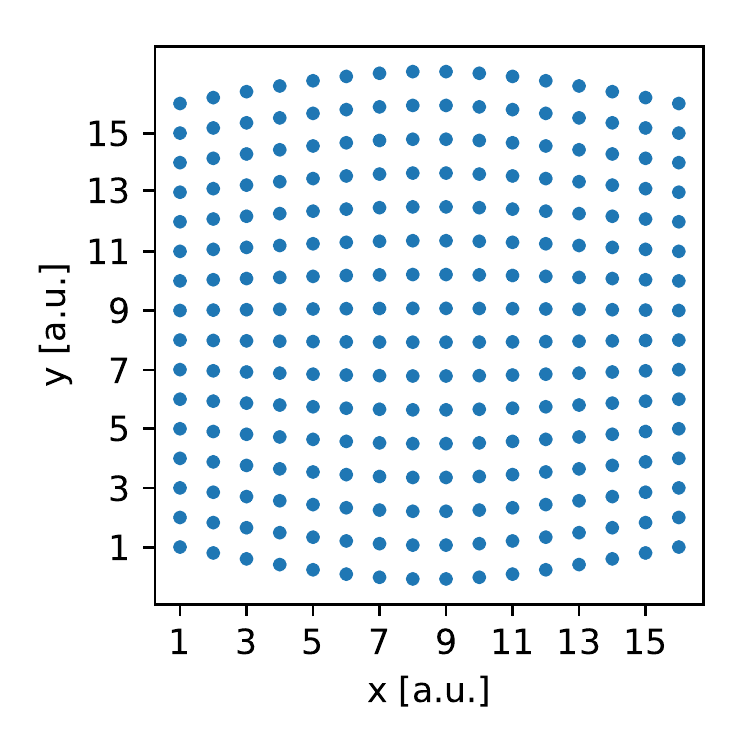}}}}
  \caption{Left: an image taken using the 4-corner readout showing the outline of the $\SI{5x5}{\cm}$ detection area using a mask of the program name of our project. The inward distortion is clearly visible. Middle: LTspice simulation of the 4-corner readout. Each point is the result of a simulation and position calculation when injecting a current into the 4-corner readout board at a pixel location. Right: By increasing the resistor values by 50\% we can compensate for the measured inward distortion in the data acquisition.}
  \label{Fig:4corner}
\end{figure}

Figure~\ref{Fig:4corner} shows an edge distortion on two opposite sides of the image (also visible in Figure~\ref{Fig:v-shape}). It also shows the result of the LTspice simulation using the currently implemented values that do not show this behaviour. However, by increasing the values of the resistor $R_0, R_1,\ldots, R_7$ by 50\% in Figure~\ref{Fig:4cornerschematic}, we are able to compensate for this effect. For the simulations we used a current source to simulate the response of a single pixel and measured the voltage drop across a \SI{50}{\ohm} resistor at the end of the 4-corner readout board. We then used equation~\ref{eq:xy} to calculate the positions shown in Figure~\ref{Fig:4corner} and ran this simulation for every pixel. Measurements with a modified setup using these new resistor values are planned.

\section{DISCUSSION}
We have described the design of an alpha particle detector for an API system for the measurement of carbon distributions in soil. The detector requirements include $<\SI{1}{\mm}$ spatial resolution and $<\SI{1}{\ns}$ timing resolution. The main detector components are a YAP scintillator, a 13700-03 Hamamatsu 16 $\times$ 16 pixel PMT, a custom 4-corner readout board, and a Pixie-16 (XIA) \SI{500}{\MHz} digitizer module. First test results using a 4-corner readout scheme demonstrated a spatial resolution well below the \SI{1}{\mm} FWHM requirement. It is noteworthy that this was achieved with a sapphire vacuum window instead of a fiber optic face plate. A readout scheme with more readout channels, 16 rows and 16 columns, will be implemented and tested in the future. Its higher rate capability is likely needed to fully take advantage of the highest rate the API system can operate at, which may result in alpha detector rates of up to \SI{1e7}{\per\second}.

\section{ACKNOWLEDGMENTS}
The information, data, or work presented herein was funded by the Advanced Research Projects Agency-Energy (ARPA-E), U.S. Department of Energy, under Contract No. DEAC02-05CH11231.

The authors also would like to thank Sergio Zimmermann, Paul Hausladen, Joshua Cates, Galina Yakubova, Alexandr Kavetskiy, and William Walburton for helpful discussion and feedback, as well as, Hui Tan and Wolfgang Henning at XIA for technical support and Takeshi Katayanagi for mechanical support.

Raw data, analysis scripts, and input files for simulations for this paper are available at Zenodo \cite{OurData}.

\bibliographystyle{aipnum-cp}%
\bibliography{paper}%

\end{document}